# High Precision Measurements of Carbon Disulfide Negative Ion Mobility and Diffusion

D.P. Snowden-Ifft and J.-L. Gauvreau

1600 Campus Rd., Occidental College, Los Angeles CA 90041, USA

## Abstract

High precision measurements were made of reduced mobility, lateral and longitudinal diffusion of $CS_2^-$ negative ions in 40 Torr $CS_2$ and a 30 – 10 Torr $CS_2$ – $CF_4$ gas mixture. The reduced mobility was found to be 353.0 +/- 0.5 cm$^2$ Torr / s V in $CS_2$ and 397.4 +/- 0.7 cm$^2$ Torr / s V in the $CS_2$ – $CF_4$ gas mixture at STP. The lateral diffusion temperatures for these two gases (295 +/- 15 K and 297 +/- 6 K) were found to be in good agreement with room temperature. By contrast longitudinal diffusion temperature was found to be slightly elevated (319 +/- 10 (stat) +/- 8 (sys) K and 310 +/- 20 (stat) +/- 6 (sys) K) though given the errors, room temperature diffusion can not be ruled out. For lateral diffusion significant capture distances (0.21 +/- 0.07 mm and 0.15 +/- 0.03 mm) were measured while for longitudinal diffusion the results were not conclusive.

## Introduction

The use of negative ions to transport charge to the readout plane is a fundamentally new way[1] of using the almost 40 year old time projection chamber[2] (TPC). Ideally, in a negative ion TPC (NITPC), electrons would be rapidly captured by molecules with moderate electron affinity at the site of ionization, transported with minimum diffusion in 3 dimensions to the readout plane where they would be stripped and then undergo normal electron avalanche to produce gas gain at the readout plane. A number of authors have found molecules approaching this ideal[1,3,4,5,6]. For the past decade the Directional Recoil Identification From Tracks (DRIFT) collaboration[7] has utilized carbon disulfide, $CS_2$ , for negative ion transport in a low pressure NITPC to search for dark matter. Analysis of alpha tracks[8] in the DRIFT experiment, which is vital for understanding backgrounds, heavily relies on the numerical value of the reduced mobility. Monte Carlo simulations of background and signals rely equally heavily on the temperature of the diffusing $CS_2$ molecules. Unfortunately the best measurements of mobility and diffusion for $CS_2$ in 40 Torr $CS_2$ are not accurate enough for the current analysis and simulations. No measurements exist for DRIFT's current gas mixture of choice 30 – 10 Torr $CS_2$ – $CF_4$. The need for high precision measurements of mobility and diffusion temperatures has become acute.

The relationship between the drift velocity, $v$, and the drift field, $E$, is determined by $E/N$, where $N$ is the gas density[9]. Despite the low pressure, 40 Torr, and high drift fields, 549 V/cm[7] with an $E/N$ = 41.6 Td (1 Td = 1e-17 V cm$^2$) negative $CS_2$ ions have been found[10] to obey the "low field approximation"[9] where $v \propto E$. Further evidence for this behavior is provided in this paper. Following Knoll[11] and Pushkin[10] we define a reduced mobility, $\mu$, with,

$$v = \mu \frac{E}{p} \tag{1}$$

where $p$ is the pressure. We refer to μ as the reduced mobility to avoid confusion with other definitions of the mobility[12]. The diffusion of negative ions in the low field approximation is given by[9],

$$\sigma^2 = \frac{2kTL}{eE} \tag{2}$$

where $\sigma$ is the rms of the spread in space of the ions after diffusing for a distance $L$ at temperature $T$. The goal of this current work was to make precise measurements (1% or better on all measurement parameters) of the reduced mobility and diffusion of $CS_2^-$ in these gases.

**Experimental Apparatus**

As in[3] the experimental method utilized photocathode generated electrons to measure negative ion reduced mobility and diffusion. A TTL pulse from the computer initiated a flash from an EGG LS-1102-1 Xe flashlamp. As shown schematically in Figure 1, a photodiode attached to the casing of the flashtube monitored the flash. Light from the flash was then collimated by a 200 μm pinhole on the end of the casing, focused by a $f_{vis}$ = 20 cm fused silica lens and passed into the vacuum vessel through a fused silica window. The light, in particular the UV portion of the light, was then focused to a 200 μm spot onto the Al photocathode by a procedure described below. $CS_2$ rapidly captured the photoelectrons and then drifted them towards the MWPC. The drift field could be varied by adjusting the voltage of the photocathode supplied by a Bertan 380X HV supply through a (1+10+1) MΩ π filter (not shown) to reduce ripple. Current from the Bertan flowed through a series of twelve 10 MΩ resistors connecting the field rings spaced 1.27 cm apart along the 15.24 cm length of the field cage ensuring a uniform drift field. The outside dimensions of the field cage were roughly 30 cm by 30 cm and the light was focused, as nearly as possible, in the center of the Al photocathode. Negative $CS_2$ ions entered the MWPC through a grid of 100 μm stainless steel wires spaced 2 mm apart and oriented horizontally as shown in Figure 1. Avalanche signals were detected on vertically oriented 20 μm stainless steel wires also spaced 2 mm apart. The anode-grid spacing for this detector is 1.11 cm and the active area was roughly 23 cm by 23 cm. The gain field for the anode wires was set by adjusting the voltage of the two sets of grid wires via a Bertan 377N HV supply also conditioned by a (1+10+1) MΩ π filter. Current from the Bertan 377N supply joined current flowing down the field cage resistor chain at the cathode wires and both currents flowed through a 66 MΩ pull-up resistor. Adjusting the voltage on the photocathode with the Bertan 380X supply therefore required adjusting the voltage of the Bertan 377N supply to insure a constant voltage on the grid wires. While slightly cumbersome, this did insure uniform drift fields from the cathode all the way to the first grid plane. The anodes were held at ground potential and were grouped down to 8 lines as follows. Wires 1, 9, 17… were grouped together to form line 1. Wires 2, 10, 18… were grouped together to form line 2. And so on. Five adjacent lines were read out using Amptek A250 charge sensitive preamplifiers. The rest were grounded. The output of the aforementioned photodiode signal and the 5 Amptek preamplifiers were sent to 6 National Instruments PXI-6133 1 MHz digitizers. This system triggered on the photodiode signal and stored 20 ms of this fast data to disk for each flash. Figure 2 shows a typical set of fast data from a single flash.

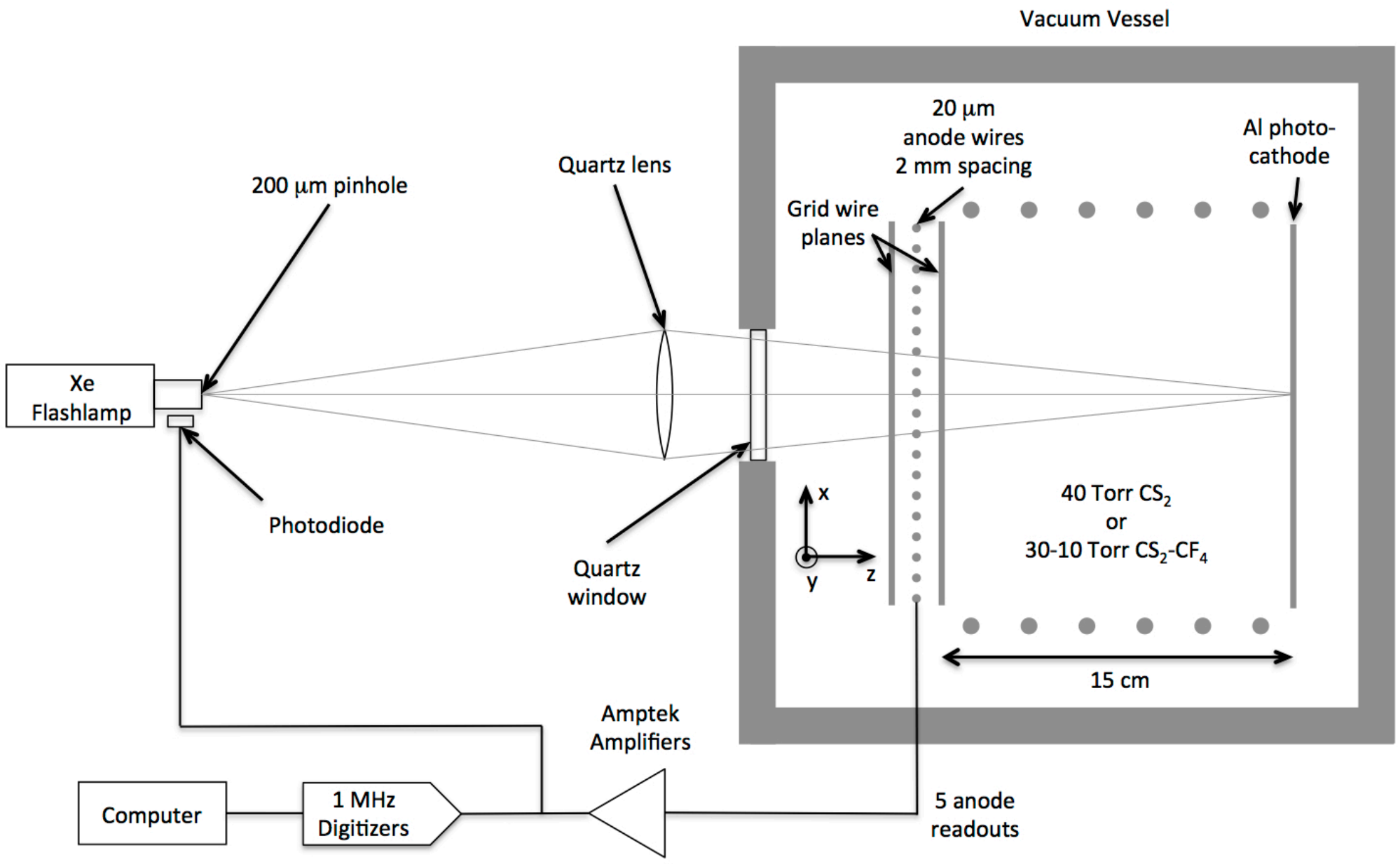

Figure 1 – Schematic of the experiment.

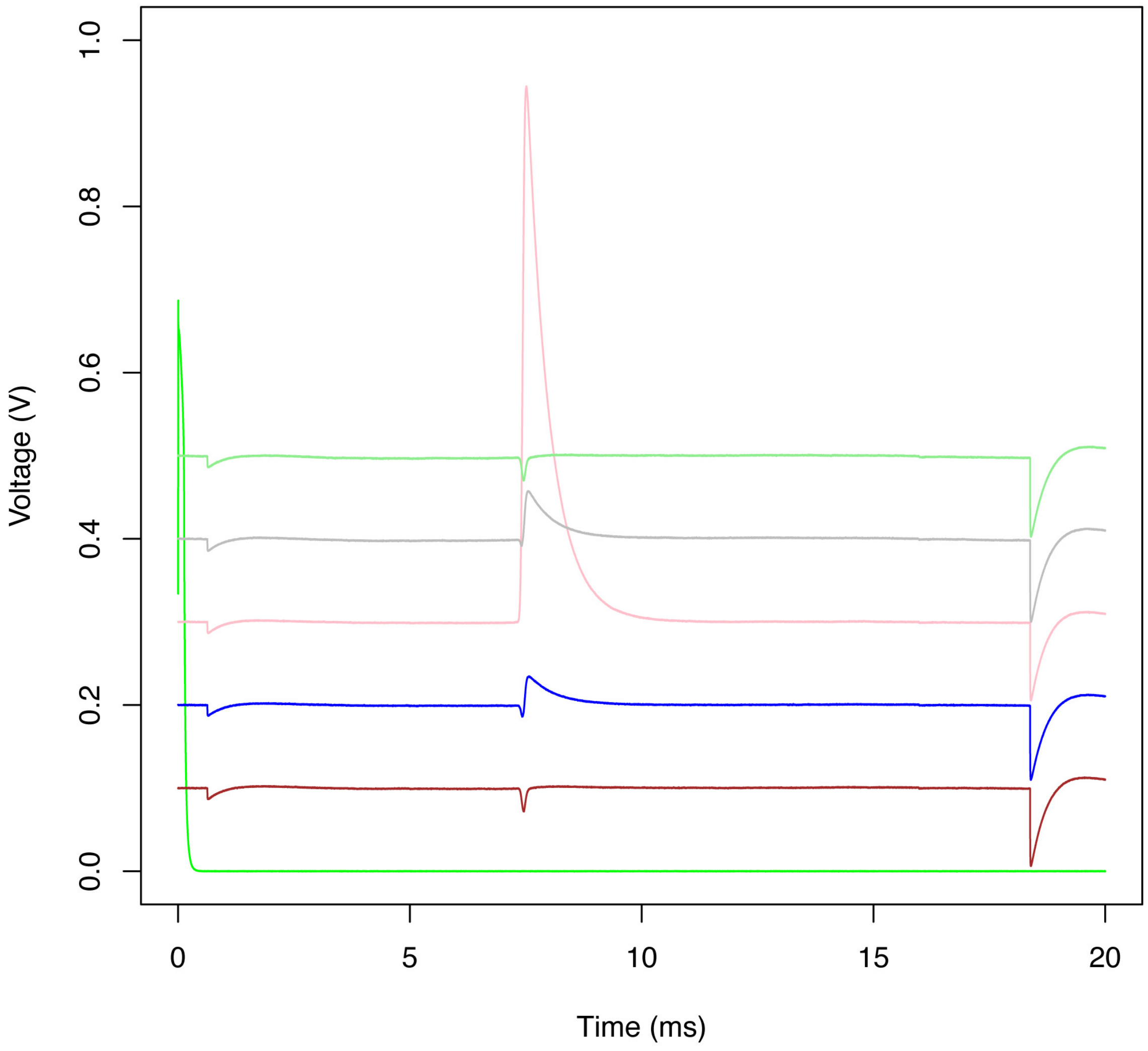

Figure 2 – "Raw" data from the fast DAQ. The baselines have been adjusted to show the events more clearly. The bottom green line is the pulse from the photodiode which triggered the system. Most of the ionization fell on the middle, pink, line for this flash at 7.4 ms. Neighboring wires show evidence of an induced pulse (negative going) with some spillover due to diffusion on the $2^{nd}$, blue, and $4^{th}$, grey lines. The excursions at roughly 0.5 ms and 18 ms are sparks.

All distances were surveyed to mm accuracy to obtain the required 1% accuracy in calculations derived from these measurements. Unfortunately the image distance of this system could not be calculated to the required accuracy because of the unknown value for the work function of the Al photocathode due to oxidation. A direct methodology was adopted instead. At relatively high drift field (~700 V/cm) with the spot centered over a single wire the flashlamp was moved toward and away from the lens until the pulse heights of neighboring wires were minimized. A microscope based visualization of the focal point in the visible revealed that the pinhole was the minimum sized object that could be focused; the Xe arc was much larger. The measured focal length of the lens, assuming the object was the pinhole, was then calculated to be 18.07 +/- 0.04 cm consistent with an average wavelength of light generating photoelectrons of 240 nm, roughly the cutoff wavelength for the combined fused silica elements.

Since the diffusion of the negative $CS_2$ ions was expected to be smaller than the 2 mm lateral spacing of the anode wires, a stepper motor was used to move the spot of light on the photocathode laterally in 200 micron increments. This was accomplished by clamping the flashlamp and lens on an optics bench which was mounted securely on an acrylic bar pivoted near the lens. The arc of the travel of this system was less than 3 degrees, insuring that the image distance remained nearly constant.

A typical scan recorded data from 80 locations, spanning 1.6 cm. Each measurement was repeated 20 times. At each of these 80 locations, slow data including the position, temperature, pressure, currents and voltages were also recorded to disk. The position of the acrylic bar was measured to sub-μm precision using a Heidenhain linear encoder. Temperatures were recorded using a thermistor calibrated with a thermocouple to 0.1 C precision and accuracy. Pressures were monitored with a Baratron 626B absolute capacitance manometer to 0.1 Torr accuracy and precision. Semiconductor grade $CF_4$ was admitted to the vacuum through a regulator. Once the proper mixture and pressure were established inside the vacuum vessel it was sealed and several scans done, typically over 8-10 hours. Leaks were monitored with the Baratron gauge overnight and we are confident that impurities were present in the 40 Torr gas mixtures to less than 0.1 Torr (0.25%). All resistances were measured to sub-% accuracy and then used to calibrate the offsets and slopes of the current and voltage readings from the two Bertan power supplies.

**E Field Calculations**

The electric field throughout the entire detector was calculated using the methodology presented in[9] for each computed grid wire and cathode voltage. As is shown in Figure 3 the grid wires were assumed to be parallel to the anode wires for this calculation. We do not believe this affects the conclusions of this paper.

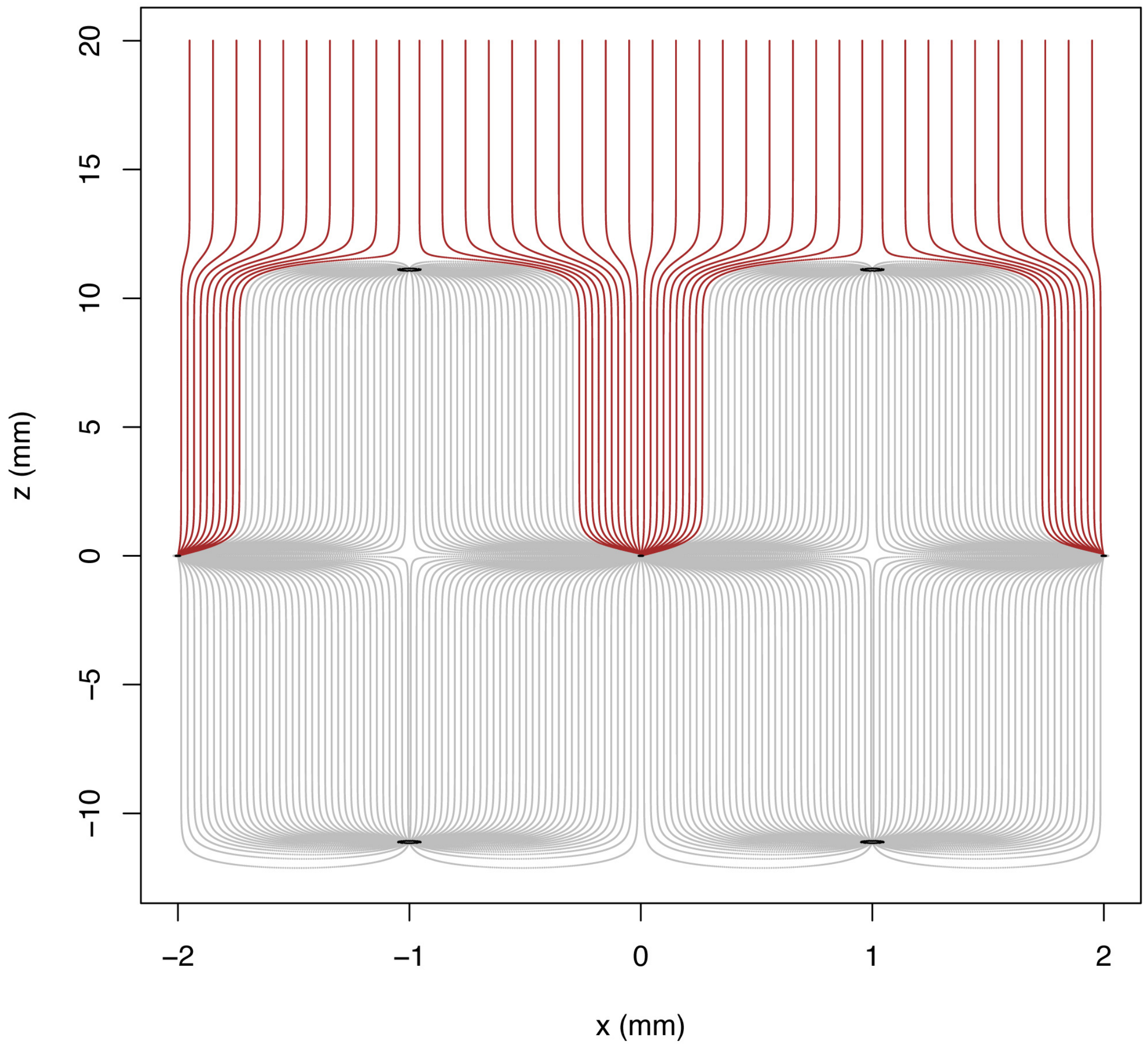

Figure 3 – Electric field lines near and in the MWPCs. Note that the horizontal dimension ($x$) is 4 mm, in total, while the vertical dimension ($z$) is ~35 mm. The 20 μm anode wires are at $z$ = 0 and $x$ = -2 mm, 0 mm and +2 mm. The grid wires are shown at $x$ = -1 mm and $x$ = +1 mm above and below the anode plane. For this simulation the grid wires are parallel to the anode wires whereas in reality they are perpendicular to the anode wires. The grey field lines are interior to the MWPC while the red field lines come into the MWPC from the drift field.

**Signal Processing**

The Amptek preamplifiers (1V/pC) integrate charge deposition on the wires with a rise time of 60 ns followed by a $\tau \sim$ 500 μS exponential decay of the voltage. A quantity proportional to the current entering the Amptek amplifiers was calculated using,

$$I(t) \propto \frac{dV}{dt} - \left(-\frac{V}{\tau}\right) \tag{3}$$

where the second term compensates for the exponential decay. Each amplifier's $\tau$ was carefully measured to the 1% level. This step was crucial in removing systematics between channels as the $\tau$s varied by 10% across the 5 fast data lines. A simple sliding average of 10 consecutive time bins (10 μS) was then applied over the entire waveform. Then a 1 ms region of interest (ROI) was established centered around the peak in this current, see Figure 4 for an example of a processed waveform. Two simple analysis cuts removed flashlamp samples in which there was a significant negative-going spark in the ROI or samples in which the baseline was significantly shifted due a prior spark and amplifier recovery.

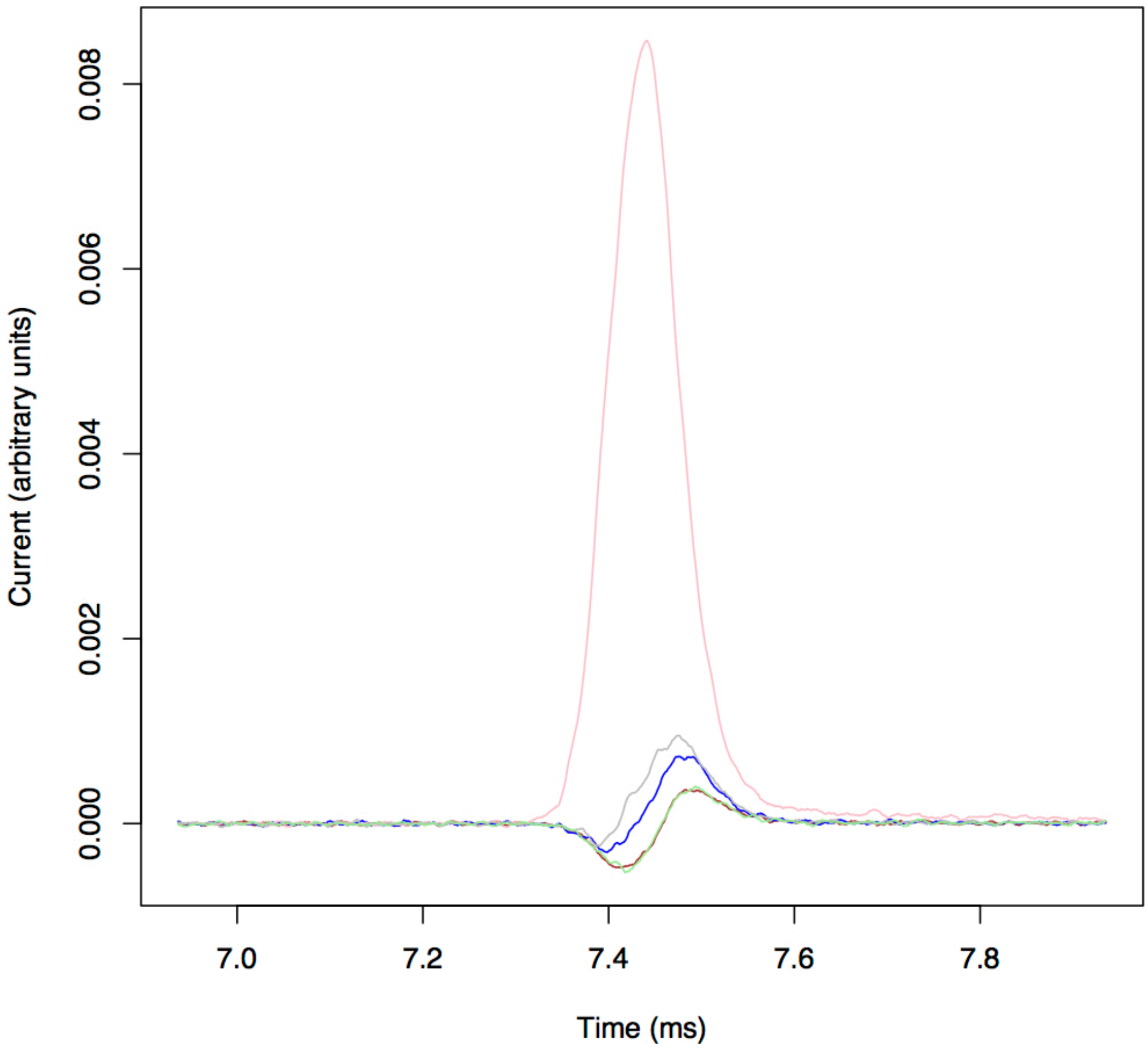

Figure 4 – Data after signal processing, described in text.

## Lateral Diffusion

For those flashlamp samples passing the analysis cuts a simple sum of the data shown in Figure 4 over the ROI provided a quantity proportional to the charge falling on that wire. Data from each line and position for one scan is shown in Figure 5. As can be seen, the data show a spread in the lateral ($x$) dimension. Unfortunately the observed spread, $\sigma_{lat,measure}$ has several contributions. Following[3],

$$\sigma^2_{lat,measure} = \frac{2kT_{lat}L}{eE} + \sigma^2_{spot} + \sigma^2_{geometry} + \sigma^2_{lat,capture} \tag{4}$$

where $\sigma_{spot}$ is the contribution from the size of the spot, $\sigma_{geometry}$ is the contribution from the 2 mm anode wire spacing and $\sigma_{lat,capture}$ is the contribution due to the capture ($e+CS_2$ -> $CS_2^-$) process. The contribution of the spot size was measured using visible light as a proxy for the UV light produced, $\sigma_{spot} = 0.101$ mm. Because the contribution of $\sigma_{geometry}$ is significant (~2mm/$\sqrt{12}$) the observed distribution of the charge was fit with a convolution of a Gaussian function and a top hat function with a 2 mm width. That fit produced an estimate of the Gaussian width of the observed distribution. $\sigma_{spot}$ was subtracted in quadrature to produce $\sigma_{lat,corrected}$,

$$\sigma^2_{lat,corrected} = \frac{2kT_{lat}}{e}\frac{L}{E} + \sigma^2_{lat,capture} \tag{5}$$

The distance shown on the horizontal axis in Figure 5 is as measured by the Heidenhain gauge and is related to the distance the spot moves by the lens magnification. The survey discussed above revealed that the magnification was 0.985. However that yielded an average distance between the wires which was systematically and significantly different from the highly accurate 2 mm wire separation distance. A ratio of 0.995 was found to provide a better fit to this distance and therefore was used for this analysis. A 1% divergence of the drift field cannot be ruled out. Data from all of the 5 lines were averaged and used to estimate the error for each measurement. Scans were taken with computed drift fields from ~100 V/cm to ~300 V/cm. According to equation (5) a plot of $\sigma_{lat,corrected}$ vs $L/E$ should yield a straight line with a slope related to the lateral diffusion temperature, $T_{lat}$, and an intercept which is related to the capture distance.

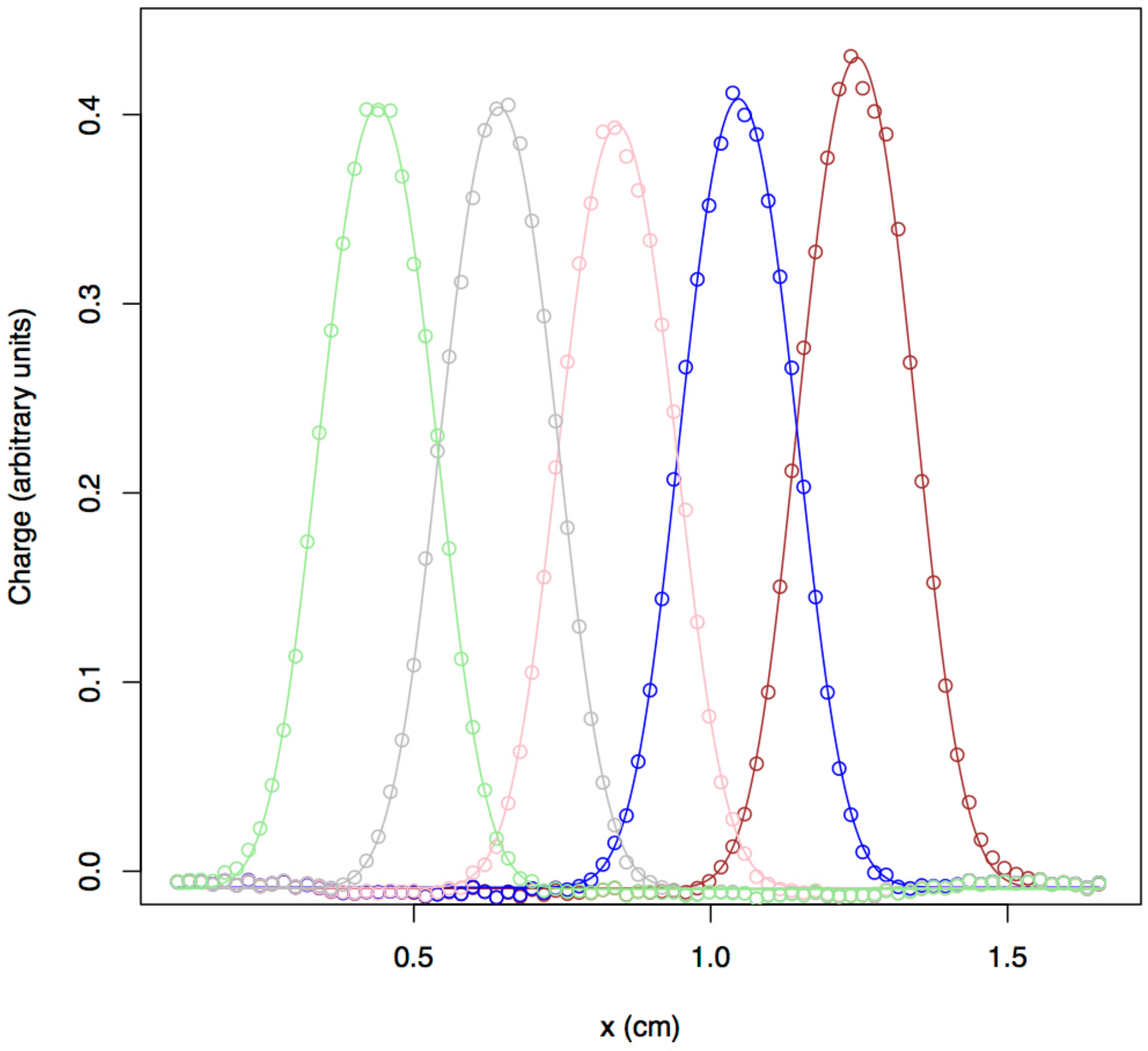

Figure 5 – Shows the lateral diffusion data from one scan. The horizontal axis is the distance measured by the linear encoder. The vertical axis is a quantity proportional to the charge falling on the wire. Each color represents a different line.

Figures 6a and 6b show plots of $\sigma^2_{lat,corrected}$ vs $L/E$ for 40 Torr $CS_2$ and the 30-10 Torr $CS_2$ – $CF_4$ gas mixture. As can be seen the data are well fit with a straight line. The extracted parameters are shown in Table 1. As can be seen both measurements indicate negative $CS_2$ ion temperatures consistent with room temperature. Room temperature diffusion temperature is evidence that the low field approximation is valid[9].

6a)

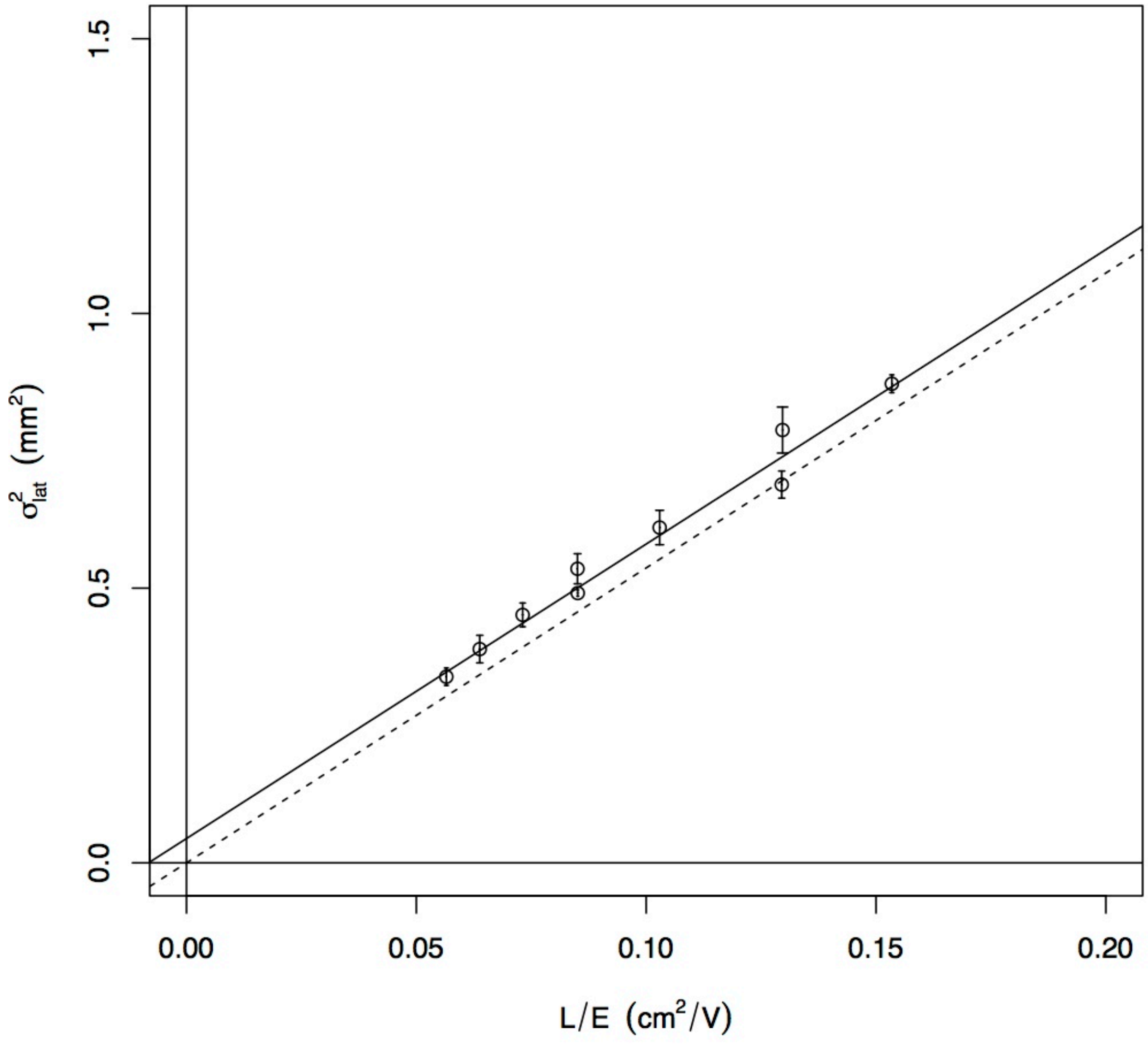

6b)

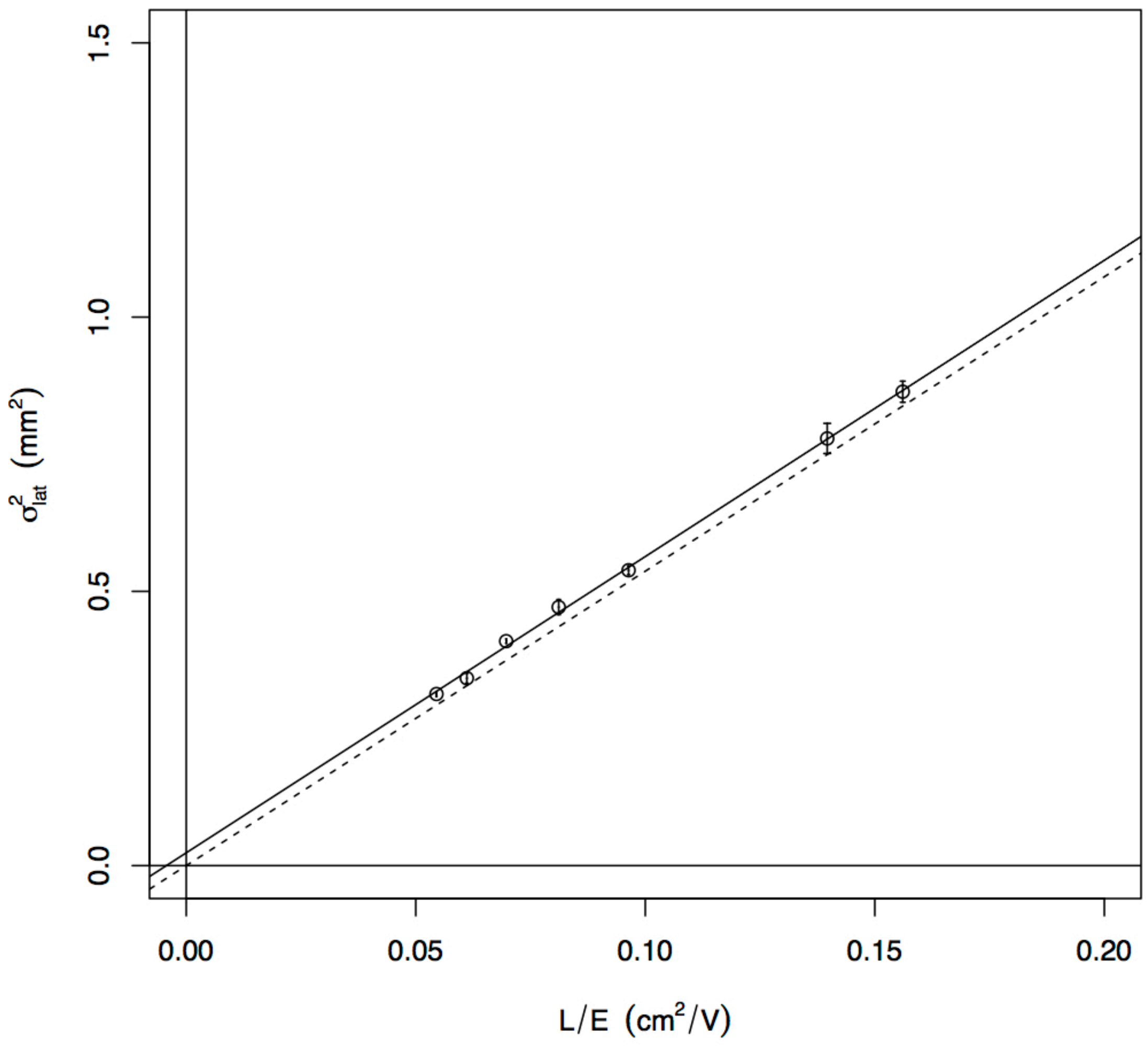

Figures 6a and 6b – Figure 6a shows lateral diffusion data for 40 Torr $CS_2$ while 6b shows data for the 30-10 Torr $CS_2$-$CF_4$ gas mixture. For both graphs the horizontal axis is *L/E* for the scan in question while the vertical axis is the observed r.m.s. width of the ionization with the spot size and 2 mm wire spacing taken into account as discussed in the text. In both cases the lower dashed line shows the ideal diffusion at room temperature with zero offset. As can be seen in both fits the observed temperature is very nearly room temperature with a finite offset.

| Table 1 – Lateral Diffusion | | |
|---|---|---|
| Gas | $T_{lat}$ (K) | $\sigma_{lat,capture}$ (mm) |
| 40 Torr $CS_2$ | 295 +/- 15 | 0.21 +/- 0.07 |
| 30-10 Torr $CS_2$ – $CF_4$ | 297 +/- 6 | 0.15 +/- 0.03 |

**Mobility**

Using the lateral scan analysis, the two closest positions to the center of a wire were located and used for mobility measurements. All flashlamp data passing the cuts were then averaged together for a particular line. Figures 7a and 7b show averaged data for line 2 at two different drift fields. The peaks of these distributions were located and used to measure the time delay, $\Delta t_{measured}$, between the flash, at $t = 0$ by definition, and the average arrival time of the ionization at the wire. The reduced mobility was then calculated using,

$$\mu = \frac{p}{\Delta t_{measured}} \int_0^L \frac{dl}{E} \tag{6}$$

where the integral was calculated using the E field model described above and the assumption that the spot was directly "over" the wire. The reduced mobility, as defined in equation (1), is temperature dependent. For instance, for equal mass drifting ions and gas molecules[9],

$$\mu = \sqrt{\frac{2kT}{3m}} \frac{e}{\sigma} \tag{7}$$

where $T$ is the temperature of the drifting ions, $m$ is their mass and $\sigma$ is the interaction cross section. For high precision measurements of the reduced mobility this temperature effect is significant. All measured reduced mobilities were therefore normalized to a STP mobility, $\mu_0$, using,

$$v = \mu_0 \sqrt{\frac{T}{T_0}} \frac{E}{p}$$

where$T_0$ = 273.15 K and the measured temperature, $T$, of the vacuum vessel during each run.

7a)

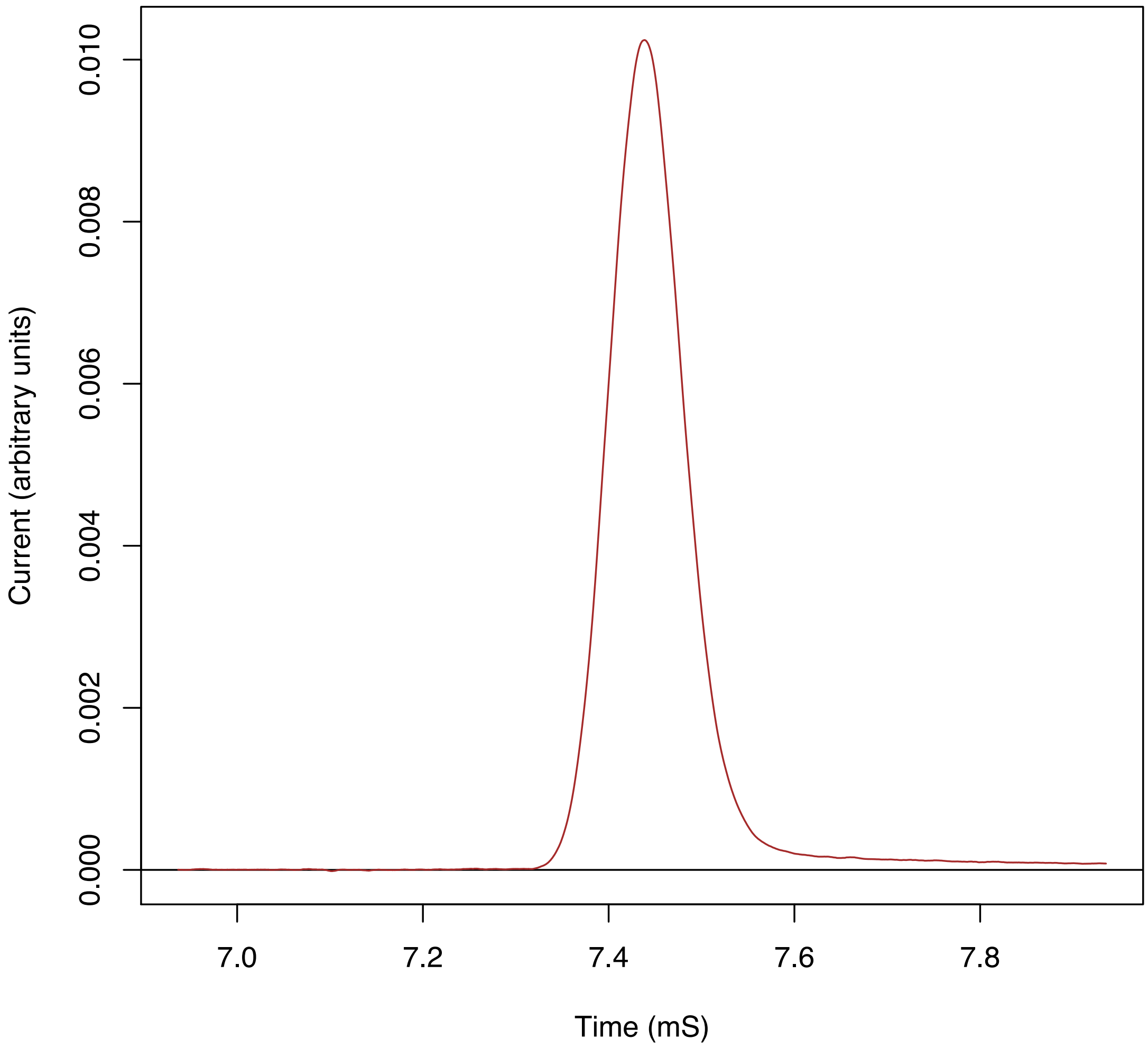

7b)

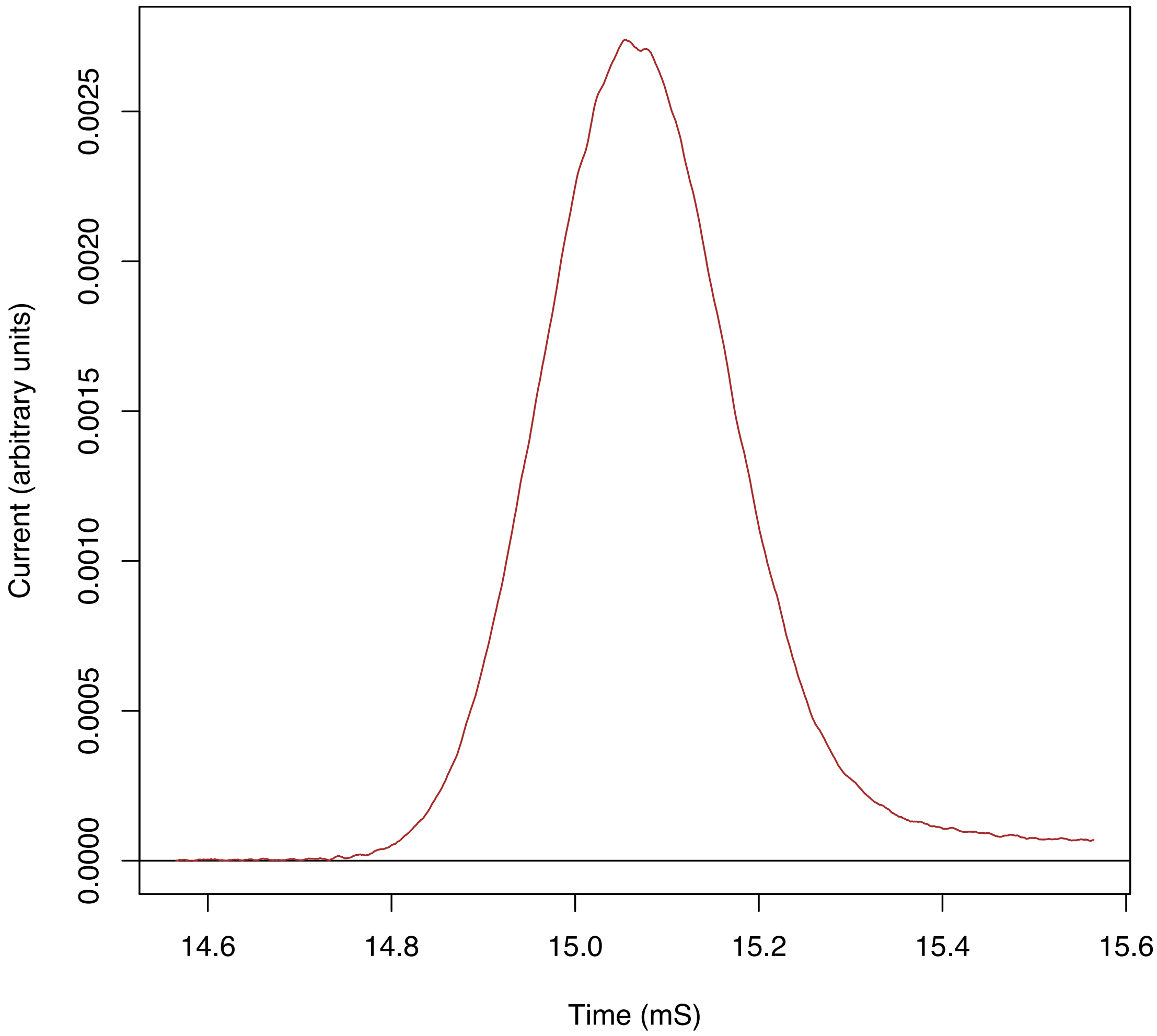

Figures 7a and 7b – Both figures show averaged waveforms for line 2 as discussed in the text. Figure 7a shows data taken at a drift field of 239 V/cm while Figure 7b shows data taken at a drift field of 118 V/cm.

Figures 8a and 8b show the obtained reduced mobility at STP as a function of the drift field *E*. As can be seen, there are variations of the measured reduced mobilities outside of statistical errors indicating the presence of some uncontrolled systematic or systematics. There is also some hint of a downward trend in the reduced mobility at very low values of *E*. The values shown in Table 2 are simple averages over all of the values shown in Figures 8a and 8b and are therefore dominated by these ~1% level systematics. The near constancy of these numbers, however, suggests that the low-field approximation is valid for this analysis.

8a)

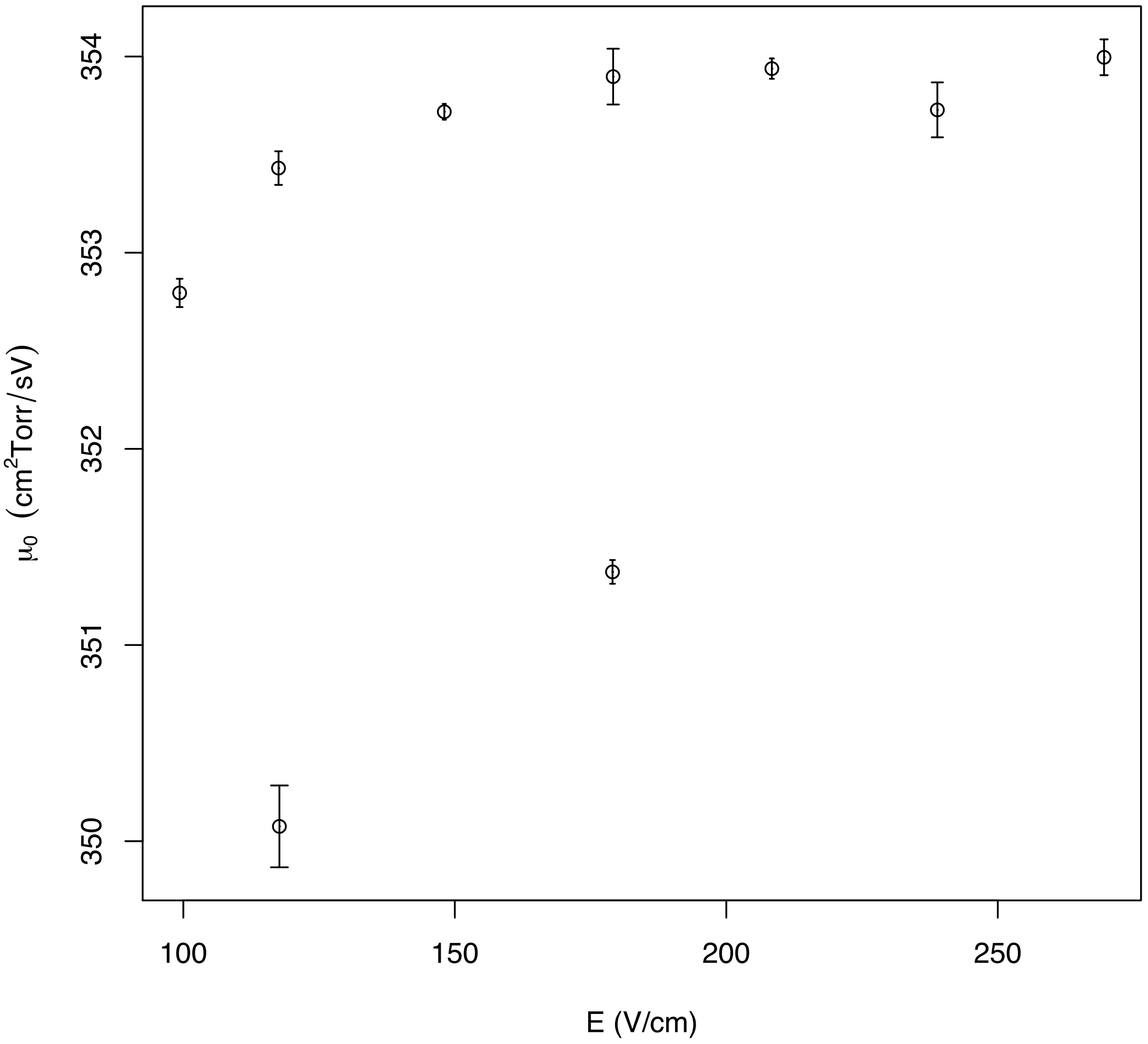

8b)

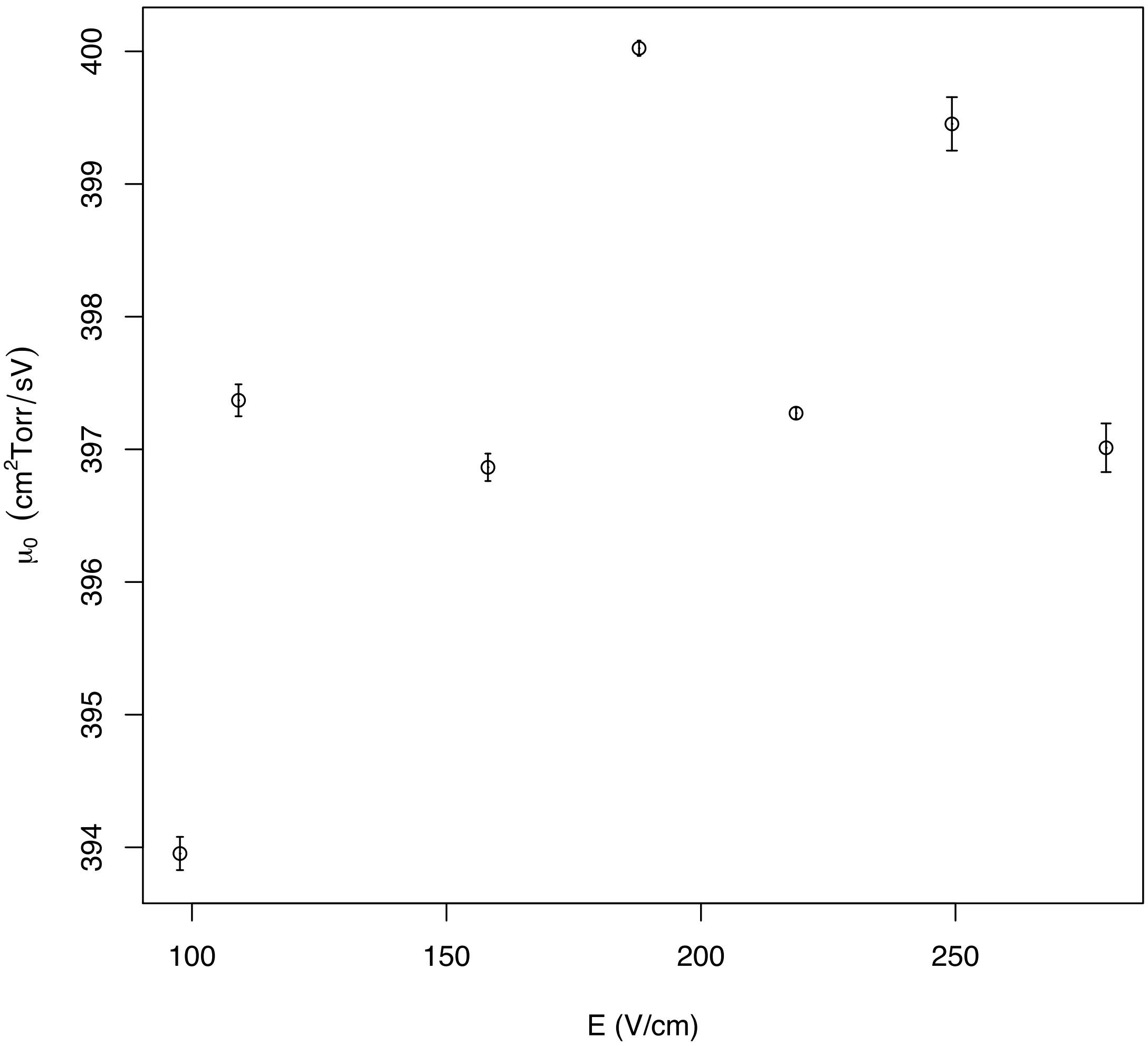

Figures 8a and 8b – Figure 8a shows the reduced mobility as a function of drift field for 40 Torr $CS_2$ normalized to STP while Figure 8b shows the same data for the 30-10 $CS_2$-$CF_4$ gas mixture. Both plots show variations well outside of random statistics indicating the presence of some uncontrolled systematic or systematics. The systematics however affect the data only at the ~1% level

| Table 2 – Reduced mobility measurements normalized to STP | |
|---|---|
| Gas | $\mu_0$ (Torr cm$^2$ / s V) @ 0 C |
| 40 Torr $CS_2$ | 353.0 +/- 0.5 |
| 30-10 Torr $CS_2$ – $CF_4$ | 397.4 +/- 0.7 |

**Longitudinal Diffusion**

Longitudinal diffusion was measured using the spread in time of the averaged waveforms, Figures 7a and 7b for instance, obtained for reduced mobility measurements and discussed above. The difference is dramatic because the width in time of the distribution depends on drift velocity as well as the diffusion, in total $\sigma_t \sim 1 / E^{3/2}$. As with the lateral diffusion measurements the measured longitudinal width, $\sigma_{long,meas}$, includes terms not associated with thermal diffusion. In this case,

$$\sigma^2_{long,meas} = \frac{2kT_{long}L}{eE} + \sigma^2_{spot} + \sigma^2_{smoothing} + \sigma^2_{path} + \sigma^2_{long,capture} \tag{9}$$

As with lateral diffusion we will assume that $\sigma_{long,capture}$ is a constant. The remaining terms, with the exception of the thermal diffusion term depend on the drift velocity. Thus,

$$(\sigma^2_{t,meas} - \sigma^2_{t,spot} - \sigma^2_{t,smoothing} - \sigma^2_{t,path})v^2 \equiv \sigma^2_{long,corrected} = \frac{2kT_{long}}{e}\frac{L}{E} + \sigma^2_{long,capture} \tag{10}$$

where $\sigma_{t,spot}$ is the dispersion in time due to the length of the flash pulse, $\sigma_{t,smoothing}$ is the dispersion induced by the smoothing and $\sigma_{t,path}$ is the dispersion in time induced by the different paths taken by the ions as they enter the MWPC, see the dark red lines of Figure 3.

Though the waveforms shown in Figures 7a and 7b appear to be Gaussian shaped, in fact they are subtly not Gaussian, as demonstrated by the presence of the long tail due to drifting positive ions following the avalanche. Fitting the entire waveforms to Gaussian functions provided poor fits and results discrepant from the procedure discussed here. To avoid the effect of the drifting ions, the waveforms were fit to Gaussians over only the beginning of the waveform, from a threshold above noise to 25% of peak height. Though this used only a fraction of the available data the quality, due to averaging ~40 waveforms, was quite high and the fits were good and consistent. Using this procedure $\sigma_{t,meas}$ was obtained.

$\sigma_{t,spot}$ was measured using the photodiode, again using the visible portion of the flash as a proxy for the UV portion, and found to be ~1 μS and therefore insignificant compared to the typical 30 - 100$\mu$s line widths. The contribution of $\sigma_{t,smoothing}$ is simply $10\mu s/\sqrt{12}$ = 2.38 μS, again insignificant. Though insignificant, they were still included in the calculation.

$\sigma_{t,path}$ was found to be neither insignificant nor constant with E field and thus was treated with some care. Figure 9 shows the arrival time as a function of vertical position, *y*, upon nearing the MWPC. As can be seen ions arriving over this entire range of *y* would suffer substantial dispersion in time. However the ions are not spread over this entire range but rather over a range given by their lateral spread,

$$\frac{2kT_{lat}}{e}\frac{L}{E}+\sigma_{spot}^2+\sigma_{lat,capture}^2 \qquad (11)$$

which depends on *E*. If the lateral dispersion is small $\sigma_{t,path}$ can also heavily depend on the vertical position, *y*, of the spot center, see Figure 9. The dependence of $\sigma_{t,path}$ on *E* significantly changes the slope of the line in Figures 10a and 10b and therefore the inferred $T_{long}$. The dependence of $\sigma_{t,path}$ on *y*, which was not measured, introduces a systematic uncertainty in our measurement and displayed in Table 3. The effect of *E* and of *y* were therefore carefully modeled using a Monte Carlo technique and included in the results below. We believe neither of these effects have been appreciated fully prior to this work.

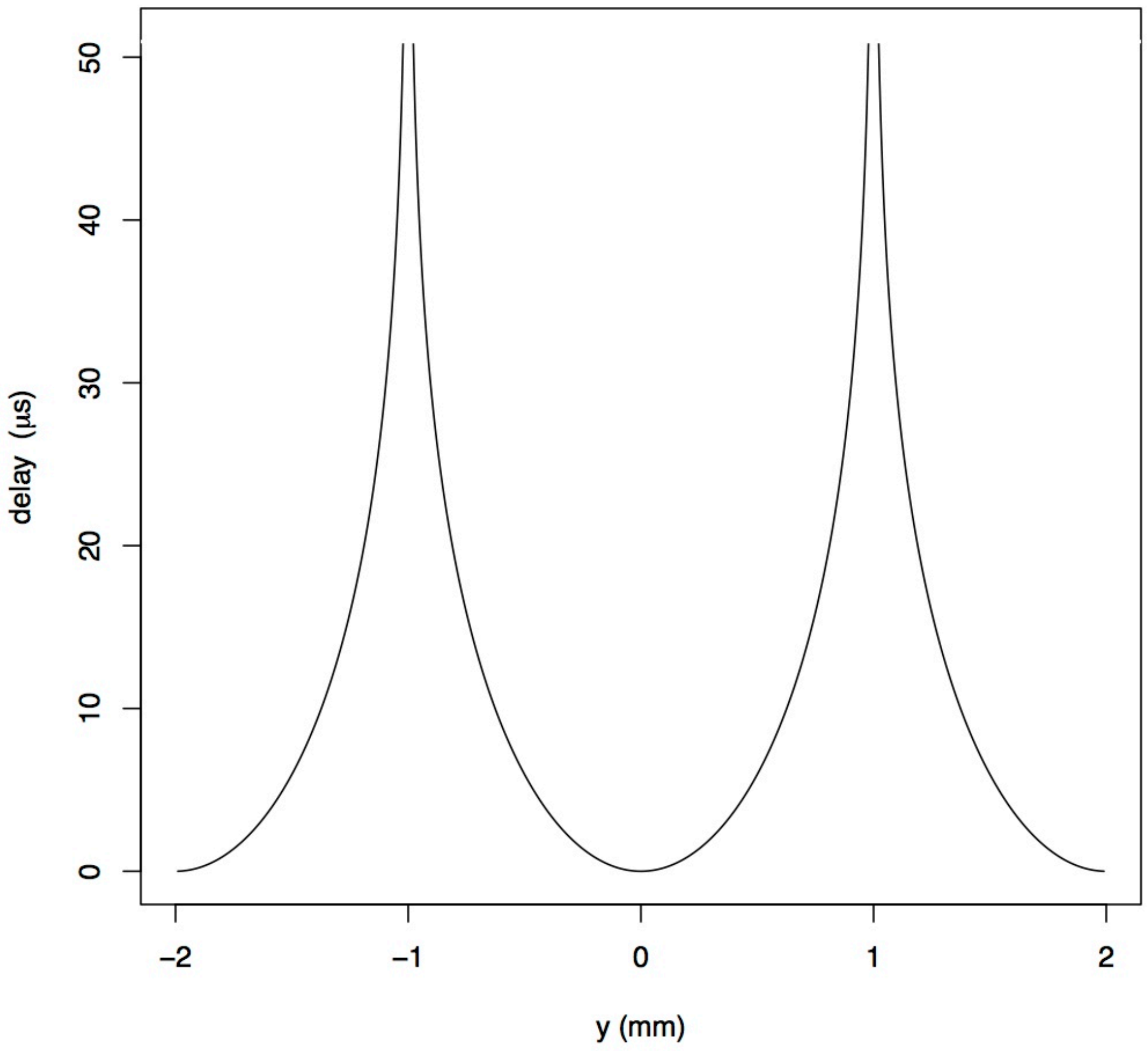

Figure 9 – The arrival time delay for ions traveling through the MWPC. The grid wires are located at y = -1 mm and +1 mm where there are two poles due to a zero in the electric field above the wires. Delay times are normalized to 0 for the shortest time, those ions falling directly between the grid wires.

Figures 10a and 10b show plots of $\sigma^2_{long,corrected}$ vs *L*/*E* for 40 Torr $CS_2$ and the 30-10 Torr $CS_2$ – $CF_4$ gas mixture. As can be seen, again, the data are well fit with a straight line. The extracted parameters are shown in Table 3. Because of the systematic effect of the *y* position of the spot only limits could be set on the capture distance.

10a)

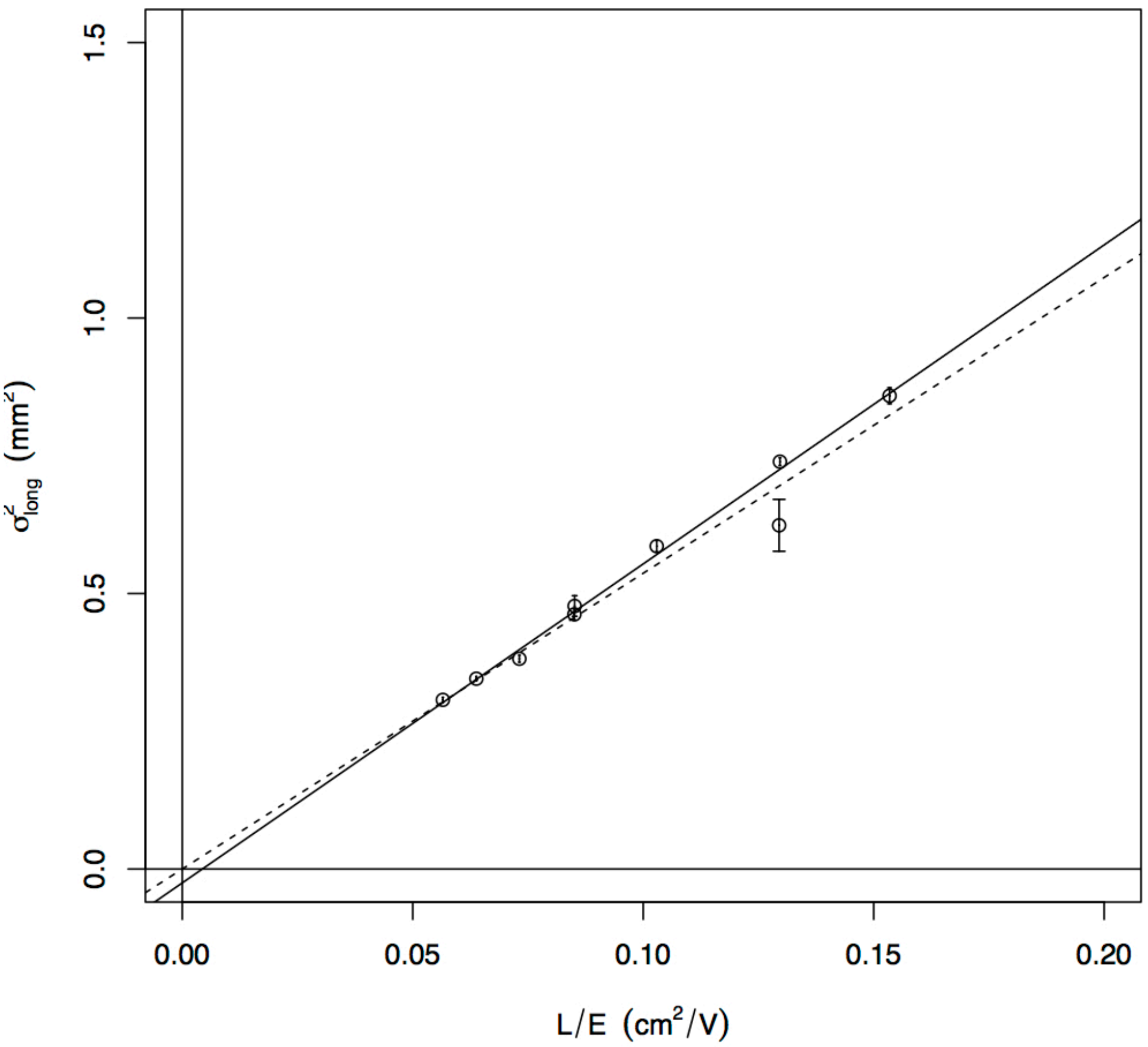

10b)

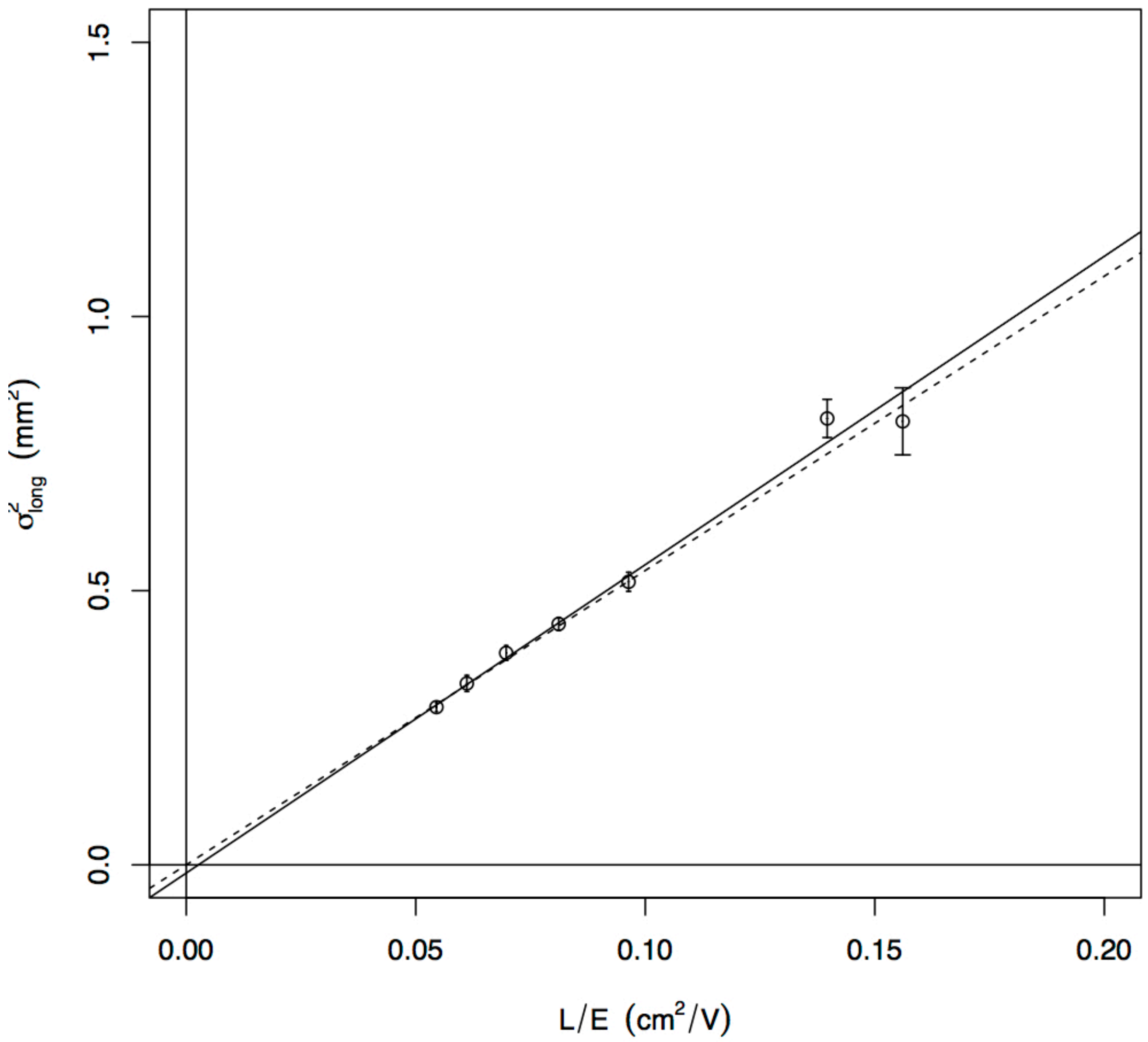

Figures 10a and 10b – Figure 10a shows longitudinal diffusion data for 40 Torr $CS_2$ while 10b shows data for the 30-10 Torr $CS_2$-$CF_4$ gas mixture. For both graphs the horizontal axis is *L/E* for the scan in question while the vertical axis is the observed r.m.s. longitudinal width of the ionization, the averaging and path length taken into account. As discussed in the text the barycenter of the spot in *y* affects this graph. For these data points *y* = 0.5 mm was assumed. In both cases the lower dashed line shows the diffusion at room temperature with zero offset. As can be seen in both fits the observed temperature is very slightly above room temperature with no offset

| Table 3 – Longitudinal Diffusion | | |
|---|---|---|
| Gas | Temperature (K) | $\sigma_{long,capture}$ (mm) |
| 40 Torr $CS_2$ | 319 +/- 10 (stat) +/- 8 (sys) | < 0.35 (90% C.L.) |
| 30-10 Torr $CS_2$ – $CF_4$ | 310 +/- 20 (stat) +/- 6 (sys) | < 0.35 (90% C.L.) |

**Discussion**

When high precision is a consideration many effects, previously not considered or ignored, come into play. Attempts were made to deal with some of these effects. More work is needed to identify the systematic associated with the reduced mobility measurements. This may be due to subtle impurities in the gas. Running the experiment with a Residual Gas Analyzer (RGA) to monitor the gas could resolve this mystery.

Several issues arose with the longitudinal measurements. The path length contribution to the measured r.m.s. in time is significant and systematic. Centering the spot halfway between two grid wires would minimize this contribution and remove this systematic. The distortions due to the drifting ions from the avalanche to the time evolution of the pulse severely restricted our ability to measure longitudinal diffusion properly. Finding the impulse response function of the MWPC and then inverting the observed waveforms to derive the arrival time of the ions at the wire would enormously improve this measurement. The addition of these last two improvements would allow for a measurement of more stringent limits on the capture distance associated with longitudinal travel and remove the systematic associated with the longitudinal temperature measurement. This is a critically important parameter for DRIFT or any experiment utilizing negative ion TPCs.

The conclusions from the lateral measurements were much more decisive. The result that the ions are diffusing at precisely room temperature is reassuring, if not unexpected. The conclusive evidence of a capture distance at sub-mm distance scales is helpful. That distance is a fundamental limiting parameter for a NITPC.

**Conclusion**

Careful measurements were made of the reduced mobility and diffusion of negative $CS_2$ anions in 40 Torr $CS_2$ and a 30-10 $CS_2$ – $CF_4$ gas mixture. These results confirm, with high precision, what previous studies of negative ion drift and diffusion concluded; that negative ion TPCs are a viable technology with numerous benefits to many fields of study. It is hoped that this research will encourage others to explore this promising technology.

## References

[1] C. J. Martoff, *et al*., Nucl. Instrum. Methods A **440**, 355 (2000).

[2] D. R. Nygren, LBL internal report, (1974).

[3] T. Ohnuki, C. J. Martoff, and D. Snowden-Ifft, Nucl. Instrum. Methods A **463,** 142 (2001).

[4] P. Sorenson, *et al*., arXiv:1205.6427.

[5] C. J. Martoff, *et al*., Nucl. Instrum. Methods A **598**, 501 (2009).

[6] M. P. Dion, *et al*., Nucl. Instrum. Methods A **648**, 186 (2011).

[7] E. Daw, *et al*., Astroparticle Physics **35**, 397 (2012).

[8] S. Burgos *et al.*, Nucl. Instrum. Methods A **584**,114 (2008).

[9] W. Blum and L. Rolandi, *Particle Detection with Drift Chambers* (Berlin, Springer, 1994).

[10] K. Pushkin and D. Snowden-Ifft, Nucl. Instrum. Methods A **606**, 569 (2009).

[11] G. Knoll, *Radiation Detection and Measurement* (New York, Wiley, 1989).

[12] Colin S. Creaser *et al.*, Analyst, 2004,**129**, 984 (2004).